# Two indicators rule them all: Mean and standard deviation used to calculate other journal indicators based on log-normal distribution of citation counts

Zhesi Shen[1], Liying Yang[1], Jinshan Wu[2]

1. National Science Library, Chinese Academy of Sciences, Beijing 100190, P. R. China
2. School of Systems Science, Beijing Normal University, Beijing, 100875, P. R. China
Email: shenzhs@mail.las.ac.cn, jinshanw@bnu.edu.cn

**Abstract**

Two journal-level indicators, respectively the mean ($m^i$) and the standard deviation ($v^i$) are proposed to be the core indicators of each journal and we show that quite several other indicators can be calculated from those two core indicators, assuming that yearly citation counts of papers in each journal follows more or less a log-normal distribution. Those other journal-level indicators include journal $h$ index, journal one-by-one-sample comparison citation success index $S_j^i$, journal multiple-sample $K^i$-$K^j$ comparison success rate $S_{j,K^j}^{i,K^i}$, and minimum representative sizes $\kappa_j^i$ and $\kappa_i^j$, the average ranking of all papers in a journal in a set of journals($R^t$). We find that those indicators are consistent with those calculated directly using the raw citation data ($\{C^i = \{c_1^i, c_2^i, \cdots, c_{N^i}^i\}, \forall i\}$) of journals. In addition to its theoretical significance, the ability to estimate other indicators from core indicators has practical implications. This feature enables individuals who lack access to raw citation count data to utilize other indicators by simply using core indicators, which are typically easily accessible.

## 1. Introduction

It is often necessary to compare journals for various purposes, e.g., evaluating journals for subscribing, selecting journals for submission. Often such comparison require full raw citation data of all papers of each individual journals in a chosen journal set. However, as a researcher or a practitioner, it might not always possible to access the full raw data. Instead, the average received citation (m) and might also its standard deviation v, of journals are often readily available. Would it be possible to rank the journals reasonably and more or less reliably only via their average received citations and the standard deviations? Since the distribution functions of received citations of most journals have heavy tails, the mean does not have its typicality, which hold very well for narrow distributions. For example, in Normal distribution, 99.7% samples are within $m \pm 3\sigma$. Due to this lack of typicality of mean, it is not reliable at all to compare journals simply according to their means.

Many other approaches and indicators, including for example percentile ranking(Pudovikin & Garfield, 2009), the Citation Success Index (CSI)(Stringer et al., 2008; Milojevic et al. 2017), fraction of highly cited papers, have been proposed to provide more reliable method of ranking journals in this challenging situation(Waltman, 2016). In this work, we focus on a specific one of those approaches, the citation success index and its extensions. This approach and the related indicator is applicable to both narrow and fat distributions. In fact, it directly makes use of the full distribution function, rather than the mean and the standard deviation. Now, our research question becomes, would it be possible to estimate those CSI-related indicators using only the means and the standard deviations. We note that under the assumption of log-normal distribution, which often holds for many journals(Radicchi et al., 2008; Thelwall, 2016A, 2016B, 2016C), it might be possible to do so. For example, in the work(Shen et al., 2018), we derived an expression of CSI in terms of only the mean and standard deviation, assuming the log-normal distribution. Therefore, in this work, we want to find similar relations for other CSI-related indicators, and if possible even other commonly used indicators, in terms of only the mean and standard deviation.

After obtaining such expressions, we aim test their accuracy by implementing them with full raw citation data of a chosen set of journals. As we will show later, we successfully find such expressions for journal level indicators including journal $\hbar$ index, journal one-by-one-sample comparison CSI $S_j^i$, journal multiple-sample $K^i$-$K^j$ comparison success rate $S_{j,K^j}^{i,K^i}$, and minimum representative sizes $\kappa_j^i$ and $\kappa_i^j$, $R^t$ the average ranking of all papers in a journal in a set of journals. Furthermore, numerical tests show very high agreement between the estimated values and the empirically calculated values of these indicators.

Our research demonstrates that by assuming a log-normal distribution of citation counts, two fundamental journal-level indicators - the mean and the standard deviation of received citations - can be utilized to compute a variety of other journal-level indicators. Therefore, these two indicators can serve as the foundational metrics for journals that generate additional indicators.

The rest of paper is organized as follow. We will first develop the formula used to calculate other indicators based on the log-normal distribution and the parameter $\{m, v\}$. We will then compare values calculated using $\{m, v\}$ against the values of indicators calculated directly from the raw data of citation counts.

## 2. Method and data

### 2.1. Log-normal distribution

We denote the citation counts, in a predefined time window say 1 year, of a paper l in a journal i as $c_l^i$ and we order them decreasingly such that $c_1^i \geq c_2^i \geq \cdots \geq c_{N^i}^i$. We assume they follows a log-normal distribution,

$$P(c) = \frac{1}{c} e^{-\frac{(\ln c - \mu_{\ln})^2}{2\sigma_{\ln}^2}}, \tag{1}$$

which is characterized by $\mu_{\ln}$ and $\sigma_{\ln}$ and thus we are going to denote the distribution function as $\rho(c, \mu_{\ln}, \sigma_{\ln})$. Given empirical data, parameter the logrithmetic mean and standard deviation $\mu_{\ln}$ and $\sigma_{\ln}$ can be estimated via

$$\begin{aligned}\mu_{\ln} &= \frac{1}{N}\sum_{l=1}^{N} \ln c_l \triangleq \langle \ln c \rangle, \\ \sigma_{\ln} &= \sqrt{\langle (\ln c)^2 \rangle - \langle \ln c \rangle^2}.\end{aligned} \tag{2}$$

Here we introduce a short hand notation, $\langle x \rangle = \frac{1}{N}\sum_{l=1}^{N} \ln x_l$.

Given empirical data, we can also estimate the arithmetic mean and standard deviation as following,

$$\begin{aligned}m &= \langle c \rangle, \\ v &= \sqrt{\langle (c)^2 \rangle - \langle c \rangle^2}.\end{aligned} \tag{3}$$

Under the assumption that the empirical data does follow an Log-normal distribution, the two set of parameters are related to each other,

$$\begin{aligned}\mu_{\ln} &= \ln\left(\frac{m}{\sqrt{1+\left(\frac{v}{m}\right)^2}}\right), \\ \sigma_{\ln} &= \sqrt{\ln\left(1 + \left(\frac{v}{m}\right)^2\right)}.\end{aligned} \tag{4}$$

Note that this formula is derived by assuming the citation counts obey the log-normal distribution with parameter $(\mu_{\ln}, \sigma_{\ln})$.

Many previous investigations(Radicchi et al., 2008; Thelwall, 2016A, 2016B, 2016C) have pointed that log-normal distribution can used to fitting the yearly citation count of journals, especially for monodisciplinary journals. In reality, it might be the case that there are some journals whose citation counts are not following exactly the log-normal distribution, log-normal distribution can still act as a rough fitting model.

With these basic mathematics, let us find the expressions of a set of indicators in terms of m and v.

**2.2. Journal Impact Factors and H-index**

Journal Impact Factor(Garfield, 1972) of a journal i is simply $m^i$. Journal $\hbar$-index ($c_{\hbar}^i \geq \hbar$)(Hirsch, 2005), which is one of the most widely used indicators and takes both quantity and quality into consideration, is defined as the largest value $\hbar$ that there are $\hbar$ papers are cited at least $\hbar$ times,

$$\hat{h} = N \times \int_{\hat{h}+1}^{\infty} \rho(x, \mu_{\ln}(m,v), \sigma_{\ln}(m,v)) dx \tag{5}$$

where $\hat{\rho}(x, \mu, \sigma)$ is the log-normal probability density function. It becomes an equation of $\hat{h}$ in terms of m and v and given m, v, then value of solution $\hat{h}$ can be solved numerically.

## 2.3. Journal one-one comparison CSI

Journal one-one comparison CSI $S_r^t$, of the journal t when compared with the journal r is defined as the probability of the citation count of an randomly selected paper from journal t being larger than that of a random paper from journal r(Stringer et al., 2008; Milojevic et al. 2017). Under the log-normal assumption, for both journals, $z_t^i \equiv \ln(c_t^i + 1) \sim N(\mu_t, \sigma_t)$ and $z_r^i \equiv \ln(c_r^i + 1) \sim N(\mu_r, \sigma_r)$. Then the condition of $c_i^t > c_j^r$ becomes $z_i^t > z_j^r$. For two normal distribution $\rho^t \equiv N(\mu^t, \sigma^t)$ and $\rho^r \equiv N(\mu^r, \sigma^r)$, the CSI, i.e. the probability of $z_i^t > z_j^r$ is calculated as(Shen et al., 2018),

$$\begin{aligned} S_r^t &= \int_{-\infty}^{\infty} \rho^r(z^r)dz^r \int_{z^r}^{\infty} \rho^t(z^t)dz^t \\ &= \iint_{z^t - z^r > 0} N(z^r, \mu^r, \sigma^r) N(z^t, \mu^t, \sigma^t) dz^r dz^t \\ &= \int_0^{\infty} N\left(x; \mu^t - \mu^r, \sqrt{(\sigma^r)^2 + (\sigma^t)^2}\right) dx \\ &= \int_{-\frac{(\mu^t - \mu^r)}{\sqrt{(\sigma^r)^2 + (\sigma^t)^2}}}^{\infty} N(x; 0,1) dx, \end{aligned} \quad (6)$$

where we have used the fact that the combination of two Normal distributions ($x = z^r - z^t$) is still a Normal distribution.

## 2.4. Journal group-group comparison CSI

The CSI is an one-one comparison between two journals. We have generalize CSI to a group-group comparison CSI $S_{r,k_r}^{t,k_t}$(Shen et al., 2019), defined as the probability of the average citation of $k_t$ random samples from journal t ($g_{k_t}^t \triangleq \frac{1}{k_t}\sum_{j=1}^{k_t} c_j^t$) being larger than the average citation of $k_r$ random samples from journal r ($g_{k_r}^r \triangleq \frac{1}{k_r}\sum_{j=1}^{k_r} c_j^r$).

There is a mathematical statement that the sum of identical log-normal variables can be very closely described by another log-normal distribution(Asmussen & Rojas-Nandayapa, 2008). Thus, $Z = \frac{1}{k}\sum_{j=1}^{k} c_j + 1$, as $c_j + 1$ follows log-normal distribution with parameters $\mu_{\ln}$ and $\sigma_{\ln}$, Z also follows very closely a log-normal distribution with following parameters $\mu_k$ and $\sigma_k$

$$\begin{aligned} \mu_k &= \mu_{\ln}(m, v) + \frac{\sigma_{\ln}^2(m,v)}{2} - \frac{\sigma_k^2}{2} + \ln k, \\ \sigma_k^2 &= \ln\left[\frac{e^{\sigma_{\ln}^2(m,v)} - 1}{k} + 1\right]. \end{aligned} \quad (7)$$

With this result, we can make use of Eq.(6) again since we are again comparing two log-normal distributions, Thus, the probability of $g_{k_t}^t > g_{k_r}^r$ can be further estimated as:

$$S_{r,k_r}^{t,k_t} = \int_{-\frac{\mu_{k_t}^t - \mu_{k_r}^r}{\sqrt{(\sigma_{k_t}^t)^2 + (\sigma_{k_r}^r)^2}}}^{\infty} N(x; 0,1) dx, \quad (8)$$

## 2.5. Minimum representative sizes

Often when comparing two similar journals with a small difference between the means, i.e. $\mu^t - \mu^r > 0$ but very small, the one-one comparison CSI $S_r^t$ is quite low. If one still want to compare the two journals, then one turn to use the group-group comparison CSI $S_{r,k_r}^{t,k_t}$. Usually, the larger the sample size $k_t, k_r$ the larger $S_{r,k_r}^{t,k_t}$. However, till how large that we may regard such comparison to be reliable? That is the definition of the minimum representative sizes $\kappa_j^i$ (Shen et al., 2019). It is defined to be the minimum value of $k_t$ and $k_r$, denoted as $\kappa_t$ and $\kappa_r$, such that $S_{r,\kappa_r}^{t,\kappa_t} \geq P_{threshold}$. The threshold value is taken for example $P_{threshold} = 0.9$. Following the same setting used in (Shen et al., 2019), that $\frac{k_r}{k_t} = \frac{v_r}{v_t}$ such that $k_r$ and $k_t$ are related so that Eq.(8) depends on only one valuable, say $k_r$. Then the the minimum representative size $\kappa_r^t$ of journal t and r for a given $P_{threshold}$ satisfies

$$S_{r,\frac{v_t}{v_r}\kappa_r^t}^{t,\kappa_r^t} = 0.9 \rightarrow \kappa_r^t, \tag{9}$$

which is an equation of $\kappa_r^t$, in terms of given values of $(m^t, v^t, m^r, v^r)$ and can be solved numerically to get $\kappa_r^t$.

**2.6. Average rank of all papers of a journal in a set of journals**

Given a set of journals, let us rank all papers according to their received citations. Then the average rank (in terms of percentile) of a journal is the average rank of all papers from this journal. Average rank of papers are often used in journal evaluation. We now provide an estimator of this average rank using also only the means and standard deviations of all journals in the set. That is,

$$R^t = \frac{1}{\sum_s N_s} \sum_r N_r S_r^t \tag{10}$$

where $N_r$ is the number of publications in journal r and $S_r^t$ is the CSI, which can be estimated simply using $m, v$, and thus also $R^t$.

Let us use the case of comparing two journals as an example. The meaning of CSI $S_r^t$ in this case, is on average, how many papers in journal t have larger received citations than those in journal r and this average is calculated exactly by ranking each paper in the two journals and then taking an average of all those ranks of papers in t. Therefore, this is exactly the meaning of average rank. Of course, it is subtle that one should note that by ranking all the papers, the papers of journal t themselves are also ordered and taken into account in this $R^t$. That is why in the right-hand side of Eq.(10), the summation r runs over the whole set including t itself.

**2.7. Data**

To implement and test our idea, papers published in 2015 and 2016 and their corresponding citation counts in 2017 from the top 30 journals listed in Web of Science-MEDICINE, GENERAL & INTERNAL category from the Journal Citation Report 2017 are extracted. Names, mean citation m and standard deviation v, and their logarithmic counterpart $\mu$ and $\sigma$ for each journal are listed in Table 1.

**Table 1. The details of the selected journals.** For each journal, its name, number of papers (N), arithmetic

mean (m = ⟨c + 1⟩) and standard deviation v, logarithmic mean (μ = ⟨ln(c + 1)⟩) and standard deviation (σ).

| ID | Journals | $N$ | $m$ | $v$ | $\mu$ | $\sigma$ |
|---|---|---|---|---|---|---|
| 1 | NEW ENGL J MED | 670 | 65.91 | 107.38 | 3.32 | 1.48 |
| 2 | LANCET | 645 | 45.02 | 63.35 | 3.32 | 0.97 |
| 3 | JAMA-J AM MED ASSOC | 410 | 36.57 | 54.68 | 3.14 | 0.93 |
| 4 | ANN INTERN MED | 302 | 17.88 | 22.40 | 2.46 | 0.89 |
| 5 | JAMA INTERN MED | 275 | 15.90 | 14.59 | 2.39 | 0.89 |
| 6 | NAT REV DIS PRIMERS | 84 | 15.88 | 12.84 | 2.50 | 0.74 |
| 7 | BMJ-BRIT MED J | 443 | 12.48 | 20.11 | 1.99 | 1.01 |
| 8 | PLOS MED | 286 | 10.94 | 11.72 | 2.02 | 0.85 |
| 9 | J CACHEXIA SARCOPENI | 88 | 9.78 | 7.02 | 2.03 | 0.75 |
| 10 | BMC MED | 398 | 8.83 | 8.53 | 1.83 | 0.85 |
| 11 | J INTERN MED | 195 | 7.34 | 7.36 | 1.65 | 0.83 |
| 12 | MAYO CLIN PROC | 282 | 7.25 | 10.81 | 1.53 | 0.90 |
| 13 | J CLIN MED | 235 | 6.29 | 6.97 | 1.44 | 0.88 |
| 14 | CAN MED ASSOC J | 143 | 5.65 | 4.43 | 1.45 | 0.77 |
| 15 | TRANSL RES | 250 | 5.40 | 4.85 | 1.36 | 0.82 |
| 16 | AM J MED | 401 | 5.22 | 5.67 | 1.27 | 0.85 |
| 17 | ANN FAM MED | 126 | 4.76 | 4.26 | 1.24 | 0.80 |
| 18 | DTSCH ARZTEBL INT | 191 | 4.68 | 3.52 | 1.30 | 0.71 |
| 19 | AMYLOID | 63 | 4.67 | 6.37 | 1.10 | 0.88 |
| 20 | AM J PREV MED | 573 | 4.61 | 5.08 | 1.17 | 0.82 |
| 21 | J GEN INTERN MED | 388 | 4.45 | 3.94 | 1.19 | 0.77 |
| 22 | PREV MED | 615 | 4.23 | 4.00 | 1.14 | 0.76 |
| 23 | PALLIATIVE MED | 182 | 4.23 | 3.14 | 1.20 | 0.71 |
| 24 | J PAIN SYMPTOM MANAG | 406 | 4.09 | 4.03 | 1.10 | 0.76 |
| 25 | AM J CHINESE MED | 192 | 4.05 | 2.85 | 1.17 | 0.69 |
| 26 | BRIT MED BULL | 101 | 4.03 | 3.33 | 1.10 | 0.76 |
| 27 | COCHRANE DB SYST REV | 1764 | 4.01 | 4.35 | 1.02 | 0.82 |
| 28 | EUR J CLIN INVEST | 255 | 3.89 | 3.05 | 1.09 | 0.74 |
| 29 | EUR J INTERN MED | 301 | 3.83 | 3.94 | 1.02 | 0.77 |
| 30 | ANN MED | 152 | 3.78 | 3.25 | 1.01 | 0.80 |

## 3. Results

In this section we will show how well the values estimated from the two basic indicators using the analytical expressions agree with the empirically calculated values of the above indicators.

### 3.1. H-index

With empirically calculated values of (m, v), we apply Eq.(5) to estimate the H-index and we then compare the estimated results against the empirically calculated H-index for all journals in our set. Note that latter need the full data of received citations of each paper in each journal. As we can see from Fig.(1), the two values show reasonable agreement.

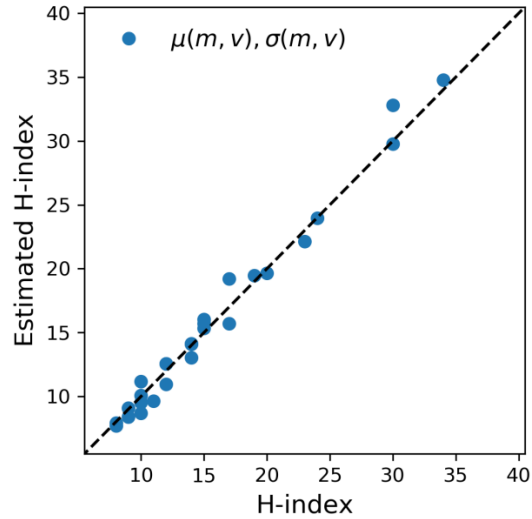

**Figure 1. Comparison of real H-index and estimated H-index.** The estimation based on m, v using Eq.(4) and Eq.(5).

### 3.2. Citation Success Index(CSI)

With empirically calculated values of $(m, v)$, we apply Eq.(6) to estimate the one-one comparison CSI $S_r^t$ and we then compare the estimated results against the empirically calculated CSI for all journals in our set. As we can see from Fig. (2), the two values show very high agreement.

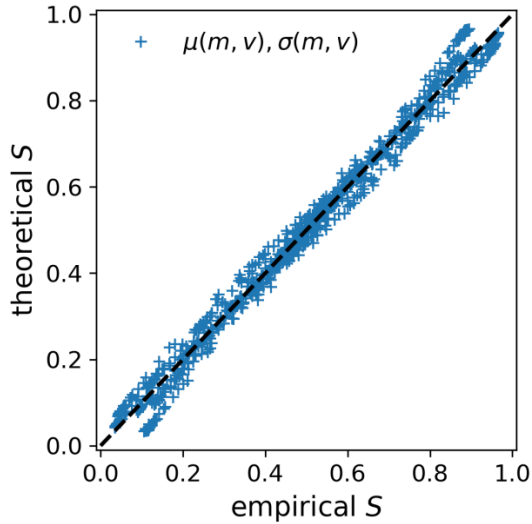

**Figure 2. Comparison of real CSI and estimated CSI.** The estimation is based on $m, v$ using Eq.(4) and Eq.(6).

### 3.3. Group-group comparison CSI

With empirically calculated values of $(m, v)$, we apply Eq.(8) to estimate the group-group comparison CSI $S_{r,k_r}^{t,k_t}$ and we then compare the estimated results against the empirically calculated group-group comparison CSI for all journals in our set. As we can see from Fig. (3), the two values show extremely high agreement.

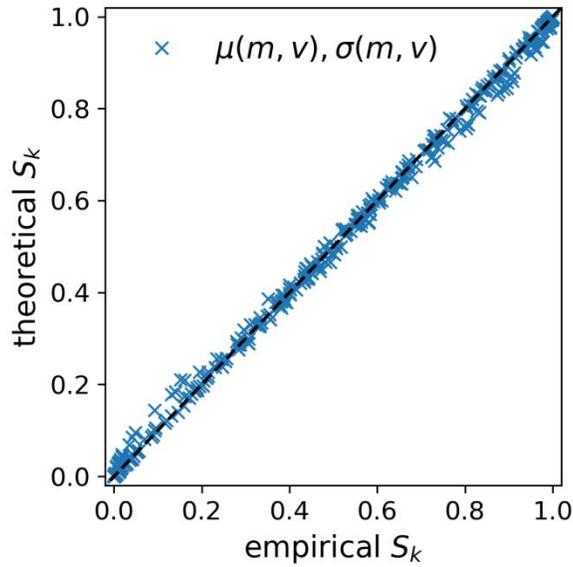

**Figure 3. Comparison of real CSI and estimated CSI at $k = 10$.** The estimation is based on m, v using Eq.(4) and Eq.(8).

### 3.4. Minimum representative number

With empirically calculated values of $(m, v)$, we apply Eq.(9) to estimate the minimum representative number $\kappa_j^i$ and we then compare the estimated results against the empirically calculated minimum representative number for all pairs of journals in our set. As we can see from Fig.(4), the two values show amazingly high agreement.

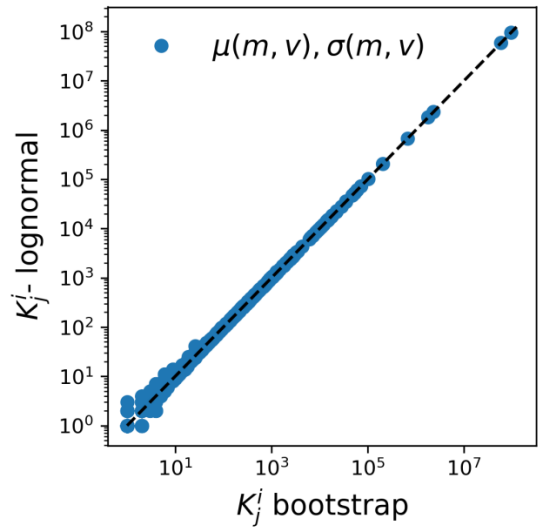

**Figure 4. Comparison of real κ and estimated κ.**

### 3.5. Average rank of all papers of a journal in a set of journals

With empirically calculated values of $(m, v)$, we apply Eq.(10) to estimate $R^i$, the average rank of all papers of a journal i and we then compare the estimated results against the empirically calculated $R^i$ for all journals in our set. As we can see from Fig.(5), the two values show reasonably high agreement.

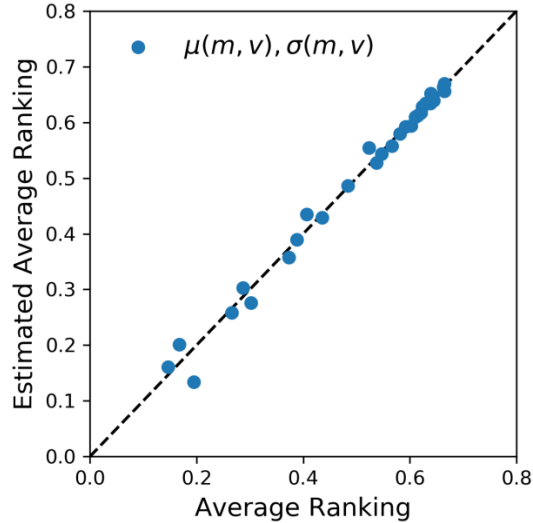

**Figure 5. Comparison of real average ranking and estimated average ranking.** The estimation is based on $m, v$ using Eq.(4) and Eq.(10).

Despite estimating these journal-level indicators using the theoretical formula derived based on the assumption of log-normal distribution, we can also calculate the more complicated indicators directly via Monte Carlo simulation: For each journal i, using the two core empirically calculated indicators $(m^i, v^i)$ as the parameter of the distribution function, we can generate $N_i$ random samples from its corresponding log-normal distribution, and then calculate these more complicated indicators. This approach is even more general. Since it does not require to have an analytic formula between the core indicators $(m^i, v^i)$ and the more complicated indicators, it can be applied to any indicators in principle. We have done so for each of the above indicators, we found very high consistency between the indicators estimated using the formula and the Monte Carlo method. How reliable can this Monte Carlo estimator be for other more complicated indicators will be the topic of further investigation.

**Conclusion and discussion**

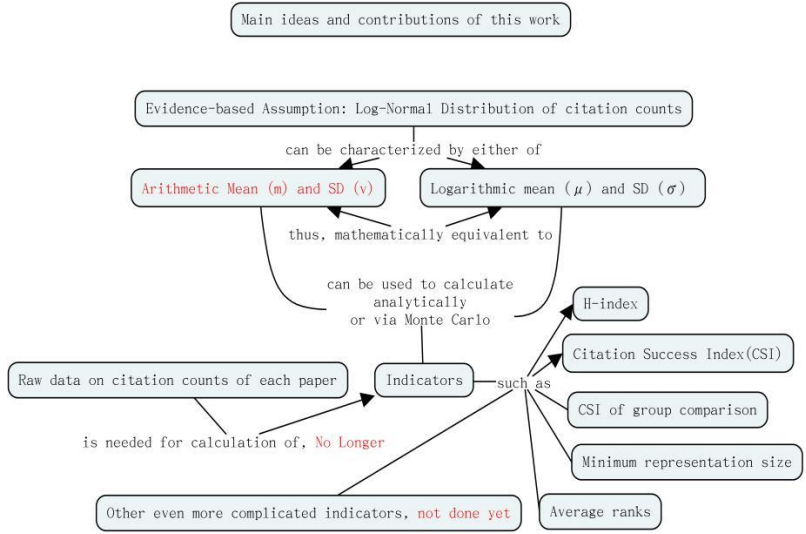

**Figure 6. A concept map illustrating the main idea and main contributions of this work.**

As summarized in Fig.(6), we have seen that with the help of log-normal distribution, journal-level indicators $\{(\mu^i, \sigma^i)\}$ are good core indicators which can be used to estimate other indicators, including H-index $h^i$, journal one-one comparison citation success index $S^i_j$, journal group-group comparison citation success index $S^{i,K^i}_{j,K^j}$, minimum representative size when comparing two journals $\kappa^i_j$, and average journal ranking $R^i$.

This possibility of estimating other indicators from two core indicators $\{(\mu^i, \sigma^i)\}$ is not to say that we should stop calculating those indicators from the raw data of citation counts. The main conclusion of this work is to show that once we know the core indicators many other indicators can be estimated, thus in a sense, providing the core indicators is more important than providing the rest. If one wants more accurate values of the indicators, then one should still calculate the indicators starting from the raw data of citation counts of each journal. Practically, for those who still need to use or calculate those indicators but without access to the raw data of citation counts, then it is possible for them to estimate those indicators using simply the core indicators.

This also does not imply that journal-level indicators should be used in evaluating individual papers. However, in the case of $\kappa^i_j = \kappa^j_i \approx 1$, which means that even taking one paper from each journal, it is very likely ($\Pr = 0.9$) that a paper from one journal has much larger number of citation counts than a paper from the other journal, then it is more or less reliable to compare papers using journal level indicators. Of course, then since all the above indicators are simply based on citation counts, another issue arises that whether or not a simply citation count can be used as good proxy of qualities in some sense of papers and journals. We do not intend to answer that question in this work.

All of our derivation and thus the estimations are based on the assumption that citation counts of papers in each individual journals follow log-normal distribution. This might not be true for some journals. That is the limitation of applicability of the proposed estimators in this work.